\documentclass[preprint,aps,12pt,showpacs,nofootinbib,tightenlines]{revtex4}
\usepackage{amsmath}
\usepackage{amssymb}
\usepackage{epsfig}
\usepackage{graphicx}
\usepackage{subfigure}
\begin{document}
%\preprint{NJNU-TH-07-05}
%%%%%%%%%%%%%%%%%%%%%%%%%%%%%%%%%%%%%%%%%%%%%
%%%%%%%%%%%%%%%%%%%%%%%%%%%%%%%%%%%%%%%%%%%%%%%%%%%%%%
\def\pslash{\rlap{\hspace{0.02cm}/}{p}}
\def\eslash{\rlap{\hspace{0.02cm}/}{e}}
%%%%%%%%%%%%%%%%%%%%%%%%%%%%%%%%%%%%%%%%%%%%%%%%%%%%%%
\title {Top quark pair production via $e^{+}e^{-}$ collision in the littlest Higgs model with T-parity at the ILC }
\author{ Bingfang Yang$^1$$^,$$^2$}\email{yangbingfang@gmail.com}
\author{ Jinzhong Han$^1$}
\author{ Lin Wang$^1$}
\author{ Xuelei Wang$^1$}\email{wangxuelei@sina.com}
\affiliation{  $^1$College of Physics $\&$ Information Engineering,
Henan Normal University, Xinxiang 453007, China\\ $^2$ Basic
Teaching Department, Jiaozuo University, Jiaozuo 454000, China
   \vspace*{1.5cm}  }

%\date{\today}
\begin{abstract}
In the littlest Higgs model with T-parity, we studied the
contributions of the new particles to the top-quark pair production
via $e^{+}e^{-}$ collision at the International Linear Collider. We
calculated the top-quark pair production cross section and found
this process can generate significantly relative correction. The
result may be a sensitive probe of the littlest Higgs model with
T-parity.
\end{abstract}
 \maketitle
%%%%%%%%%%%%%%%%%%%%%%%%%%%%%%%%%%%%%%%%%%%%%%%%%%%%%%%%%%%%%%%%%%%%%%%%%%%%%%%
\section{ Introduction}
\noindent Since the top quark was first observed at Ferminlab
Tevatron in 1995 \cite{1}, it is always one of the forefront topics
and being studied heatedly. The top quark is the heaviest elementary
fermion we have discovered and its mass is very close to the
electroweak scale, so it may play an important role in the mechanism
of electroweak symmetry breaking (EWSB) \cite{2} and is considered
as an ideal tool for probing new physics (NP) beyond the standard
model (SM). The top-quark properties have not been precisely
measured at the Tevatron due to the small statistics. But about
$10^{7}$ top pair signals per year will be produced at the LHC,
which will be suitable to determine the top-quark properties. On the
other hand, although the cross section for $t\bar{t}$ production at
the International Linear Collider(ILC) \cite{3} is less than the
LHC, the ILC will be an ideal place for further and complementary
investigation of top quark compared to the complicated QCD
background of the LHC as well as polarization of the initial beams.

The little Higgs theory was proposed \cite{4} as a possible solution
to the hierarchy problem and always remains a popular candidate for
the NP. As a cute economical implementation of the little Higgs, the
littlest Higgs (LH) model \cite{5} is suffered from severe
constraints from electroweak precision tests \cite{6} so that the
fine-tuning was reintroduced in the Higgs potential \cite{7}. Due to
the exchanges of additional heavy gauge bosons in the theories, the
most serious constraints resulted from the tree-level corrections to
precision electroweak observables, as well as from the small but
non-vanishing vacuum expectation value (VEV) of an additional
weak-triplet scalar field. In order to solve this problem, a
discrete symmetry called T-parity is proposed \cite{8}, which
explicitly forbids any tree-level contribution from the heavy gauge
bosons to the observables involving only SM particles as external
states. It also forbids the interactions that induce triplet VEV
contributions. This resulting model is referred to as the littlest
Higgs model with T-parity (LHT).

The LHT model predicts heavy T-odd gauge bosons which are the
T-partners of the SM gauge boson and also heavy T-odd $SU(2)$
doublet fermions. In the LHT model, the top-quark sector is quite
complicated. One aspect of its phenomenology in top-quark sector is
that there are interactions between the heavy fermions and the heavy
gauge bosons, which can contribute to the $Vt\bar{t}(V=\gamma,Z)$
couplings and give corrections to the $t\bar{t}$ production cross
section. In this paper, we study the LHT contributions to the
relative corrections of the top-quark pair production cross section
in $e^{+}e^{-}$ collision at the ILC. Because of this unique
structure in top-quark sector, the results may be utilized to probe
the LHT model.

This paper is organized as follows. In Sec.II we give a brief review
of the LHT model related to our work. In Sec.III we calculate the
tree and one-loop level contributions of the LHT model to the
$e^{+}e^{-}\rightarrow t\bar{t}$ and show some figures of this
process in the LHT model at the ILC. Finally, we give our
conclusions in Sec.IV.

\section{ A brief review of the LHT model}
 \noindent The LHT model \cite{8} is based on an $SU(5)/SO(5)$ non-linear sigma model, where the
global group $SU(5)$ is spontaneously broken into $SO(5)$ at the
scale $f\sim\mathcal{O}(TeV)$. From the $SU(5)/SO(5)$  breaking,
there arise 14 Goldstone bosons which are described by the ``pion"
matrix $\Pi$ as follows
\begin {equation}
\Pi=
\begin{pmatrix}
-\frac{\omega^0}{2}-\frac{\eta}{\sqrt{20}}&-\frac{\omega^+}{\sqrt{2}}
&-i\frac{\pi^+}{\sqrt{2}}&-i\phi^{++}&-i\frac{\phi^+}{\sqrt{2}}\\
-\frac{\omega^-}{\sqrt{2}}&\frac{\omega^0}{2}-\frac{\eta}{\sqrt{20}}
&\frac{v+h+i\pi^0}{2}&-i\frac{\phi^+}{\sqrt{2}}&\frac{-i\phi^0+\phi^P}{\sqrt{2}}\\
i\frac{\pi^-}{\sqrt{2}}&\frac{v+h-i\pi^0}{2}&\sqrt{4/5}\eta&-i\frac{\pi^+}{\sqrt{2}}&
\frac{v+h+i\pi^0}{2}\\
i\phi^{--}&i\frac{\phi^-}{\sqrt{2}}&i\frac{\pi^-}{\sqrt{2}}&
-\frac{\omega^0}{2}-\frac{\eta}{\sqrt{20}}&-\frac{\omega^-}{\sqrt{2}}\\
i\frac{\phi^-}{\sqrt{2}}&\frac{i\phi^0+\phi^P}{\sqrt{2}}&\frac{v+h-i\pi^0}{2}&-\frac{\omega^+}{\sqrt{2}}&
\frac{\omega^0}{2}-\frac{\eta}{\sqrt{20}}
\end{pmatrix}
\end{equation}
Under T-parity the SM Higgs doublet,
$H=(-i\pi^+\sqrt{2},(v+h+i\pi^0)/2)^T$ is T-even while other fields
are T-odd.

The Goldstone bosons $\omega^{\pm},\omega^{0},\eta$ are eaten by the
new T-odd gauge bosons $W_{H}^{\pm},Z_{H},A_{H}$ respectively, whose
masses up to $\mathcal O(\upsilon^{2}/f^{2})$ are given by
\begin {equation}
M_{W_{H}}=M_{Z_{H}}=gf(1-\frac{\upsilon^{2}}{8f^{2}}),M_{A_{H}}=\frac{g'f}{\sqrt{5}}
(1-\frac{5\upsilon^{2}}{8f^{2}})
\end {equation}
with $g$ and $g'$ being the SM $SU(2)$ and $U(1)$ gauge couplings,
respectively.

The Goldstone bosons $\pi^{\pm},\pi^{0}$ are eaten by the
$W^{\pm}$and $Z$ bosons of the SM, whose masses up to $\mathcal
O(\upsilon^{2}/f^{2})$ are given by
\begin {equation}
M_{W_{L}}=\frac{g\upsilon}{2}(1-\frac{\upsilon^{2}}{12f^{2}}),M_{Z_{L}}=\frac{g\upsilon}
{2\cos\theta_{W}}(1-\frac{\upsilon^{2}}{12f^{2}})
\end {equation}
The photon $A_{L}$ remains massless and is also T-even. Where ``L"
and ``H" denote ``light" and ``heavy", respectively.

In order to preserve the T-parity, for each SM fermion, a copy of
mirror fermion with T-odd quantum number is added. We denote them by
$u_{H}^{i},d_{H}^{i},l_{H}^{i},\nu_{H}^{i}$, where i= 1, 2, 3 are
the generation index. Neglecting the $\mathcal
O(\upsilon^{2}/f^{2})$ correction, the masses of the mirror fermions
are given in a unified manner:
\begin{equation}
m_{F_{H}^{i}}=\sqrt{2}\kappa_if
\end{equation}
 where $\kappa_i$ are the diagonalized Yukawa couplings of the mirror fermions.

In the top sector, an additional heavy quark $T_{+}$ is introduced
to cancel the quadratic divergence of the Higgs mass induced by top
loops. Under T-parity, $T_{+}$ is even. The implementation of
T-parity then requires a T-odd partner $T_{-}$. Their masses are
given by
\begin{eqnarray}
m_{T_{+}}&=&\frac{f}{v}\frac{m_{t}}{\sqrt{x_{L}(1-x_{L})}}[1+\frac{v^{2}}{f^{2}}(\frac{1}{3}-x_{L}(1-x_{L}))]\\
m_{T_{-}}&=&\frac{f}{v}\frac{m_{t}}{\sqrt{x_{L}}}[1+\frac{v^{2}}{f^{2}}(\frac{1}{3}-\frac{1}{2}x_{L}(1-x_{L}))]
\end{eqnarray}
where $x_{L}$ is the mixing parameter between the SM top-quark $t$
and the new top-quark $T_{+}$.

In the LHT model, the mirror fermions open up a new flavor structure
in the model. One of the important ingredients of the mirror sector
is the existence of four CKM-like unitary mixing matrices, two for
mirror quarks and two for mirror leptons:
\begin{equation}
V_{Hu},V_{Hd},V_{Hl},V_{H\nu}
\end{equation}
where $V_{Hu}$ and $V_{Hd}$ are for the mirror quarks, $V_{Hl}$ and
$V_{H\nu}$ are the mirror leptons mixing matrices. These mirror
mixing matrices are involved in the flavor changing interactions
between the SM fermions and the mirror fermions which are mediated
by the T-odd gauge bosons or T-odd Goldstone bosons. They satisfy
the relation $V_{Hu}^{\dag}V_{Hd}=V_{CKM}$ and
$V^{\dag}_{H\nu}V_{Hl}=V_{PMNS}$. We follow Ref.\cite{9} to
parameterize $V_{Hd}$ with three angles
$\theta^d_{12},\theta^d_{23},\theta^d_{13}$ and three phases
$\delta^d_{12},\delta^d_{23},\delta^d_{13}$
\begin{eqnarray}
V_{Hd}=
\begin{pmatrix}
c^d_{12}c^d_{13}&s^d_{12}c^d_{13}e^{-i\delta^d_{12}}&s^d_{13}e^{-i\delta^d_{13}}\\
-s^d_{12}c^d_{23}e^{i\delta^d_{12}}-c^d_{12}s^d_{23}s^d_{13}e^{i(\delta^d_{13}-\delta^d_{23})}&
c^d_{12}c^d_{23}-s^d_{12}s^d_{23}s^d_{13}e^{i(\delta^d_{13}-\delta^d_{12}-\delta^d_{23})}&
s^d_{23}c^d_{13}e^{-i\delta^d_{23}}\\
s^d_{12}s^d_{23}e^{i(\delta^d_{12}+\delta^d_{23})}-c^d_{12}c^d_{23}s^d_{13}e^{i\delta^d_{13}}&
-c^d_{12}s^d_{23}e^{i\delta^d_{23}}-s^d_{12}c^d_{23}s^d_{13}e^{i(\delta^d_{13}-\delta^d_{12})}&
c^d_{23}c^d_{13}
\end{pmatrix}
\end{eqnarray}
and follow Ref.\cite{10} to parameterize $V_{Hl}$ with three angles
$\theta^l_{12},\theta^l_{23},\theta^l_{13}$ and three phases
$\delta^l_{12},\delta^l_{23},\delta^l_{13}$
\begin{eqnarray}
V_{Hl}=
\begin{pmatrix}
c^l_{12}c^l_{13}&s^l_{12}c^l_{13}e^{-i\delta^l_{12}}&s^l_{13}e^{-i\delta^l_{13}}\\
-s^l_{12}c^l_{23}e^{i\delta^l_{12}}-c^l_{12}s^l_{23}s^l_{13}e^{i(\delta^l_{13}-\delta^l_{23})}&
c^l_{12}c^l_{23}-s^l_{12}s^l_{23}s^l_{13}e^{i(\delta^l_{13}-\delta^l_{12}-\delta^l_{23})}&
s^l_{23}c^l_{13}e^{-i\delta^l_{23}}\\
s^l_{12}s^l_{23}e^{i(\delta^l_{12}+\delta^l_{23})}-c^l_{12}c^l_{23}s^l_{13}e^{i\delta^l_{13}}&
-c^l_{12}s^l_{23}e^{i\delta^l_{23}}-s^l_{12}c^l_{23}s^l_{13}e^{i(\delta^l_{13}-\delta^l_{12})}&
c^l_{23}c^l_{13}
\end{pmatrix}
\end{eqnarray}
\section{Top quark pair production via $e^{+}e^{-}$ collision in the LHT model}
\noindent  The relevant Feynman diagrams for the LHT model
contributions to $e^{+}e^{-}\rightarrow t\bar{t}$ are shown in
Fig.(1-3). We can see that the Feynman diagrams consist of vertex
correction, propagator correction and box diagrams. The diagrams of
T-odd particles are induced by the interactions between the SM
fermions and the mirror fermions mediated by the heavy T-odd gauge
bosons or T-odd Goldstone bosons.

Due to the mixing between $t$ and $T_{+}$, the $Zt\bar{t}$ coupling
is modified at tree level, which can be expressed as
\begin{equation}
\frac{ig}{C_{W}}\gamma_{\mu}[(\frac{1}{2}-\frac{2}{3}S_{W}^{2}-\frac{1}{2}x_{L}^{2}\frac{v^{2}}{f^{2}})P_{L}-\frac{2}{3}S_{W}^{2}P_{R}],
\end{equation}
where $\theta_{W}$ is the Weinberg angle, $S_{W}=sin\theta_{W}$,
$C_{W}=cos\theta_{W}$, and $P_{L}=\frac{1-\gamma_{5}}{2}$ and
$P_{R}=\frac{1+\gamma_{5}}{2}$ are the left-handed and right-handed
projection operators, respectively. We can see that the strength of
the left-handed $Zt\bar{t}$ coupling in the LHT becomes smaller than
that in the SM by the term
$-\frac{1}{2}x_{L}^{2}\frac{v^{2}}{f^{2}}$ and the right-handed
$Zt\bar{t}$ coupling is the same as the SM.

 In the NLO calculation, the high order $\mathcal
O(\upsilon^{2}/f^{2})$ terms in the masses of new particles and in
the Feynman rules are both neglected, the invariant amplitudes are
considered to the order $\mathcal O(\upsilon/f)$. For this reason,
the higher order couplings and the couplings between the scalar
triplet $\Phi$ and top quark are neglected. We can calculate the
loop diagrams straightforwardly. Each loop diagram is composed of
some scalar loop functions \cite{11}, which can be calculated by
using LOOPTOOLS \cite{12}. The relevant Feynman rules can be found
in Ref.\cite{13}. We use the 't Hooft-Feynman gauge, so the
Goldstone bosons and the ghost fields should be involved. The
analytic expressions of the amplitudes for these processes are
lengthy and tedious, so we don't give the explicit expressions.

We use the dimensional regularization scheme to regulate the
ultraviolet divergences and adopt the on-shell renormalization
scheme to renormalize the electroweak parameters. We have checked
the divergences and found the divergences of the renormalized
propagator and the renormalized vertex have been canceled. There
aren't divergences in the box diagrams.

The SM parameters used in our calculations are\cite{14}
\begin{eqnarray}
\nonumber G_{F}=1.16637\times 10^{-5}GeV^{-2}, S_{W}^{2}=0.231,
\alpha_{e}=1/128,\\M_{Z_{L}}=91.2GeV,m_{t}=172.4GeV,m_{h}=120GeV.~~~~
\end{eqnarray}

The relevant LHT parameters in our calculation are the scale $f$,
the mixing parameter $x_{L}$, the mirror fermion masses and
parameters in the matrices $V_{Hu},V_{Hd}$ and $V_{Hl},V_{H\nu}$.

\begin{figure}[htbp]
\scalebox{0.4}{\epsfig{file=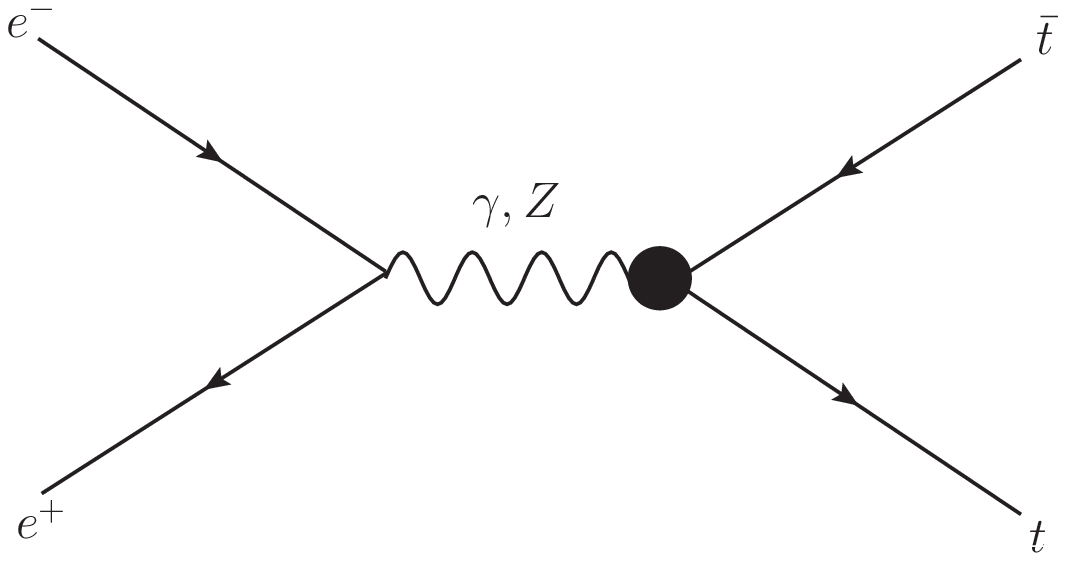}}
\end{figure}
Where the effective vertices $\gamma/Zt\bar{t}$ are shown in Fig.1.
\begin{figure}[htbp]
\scalebox{0.4}{\epsfig{file=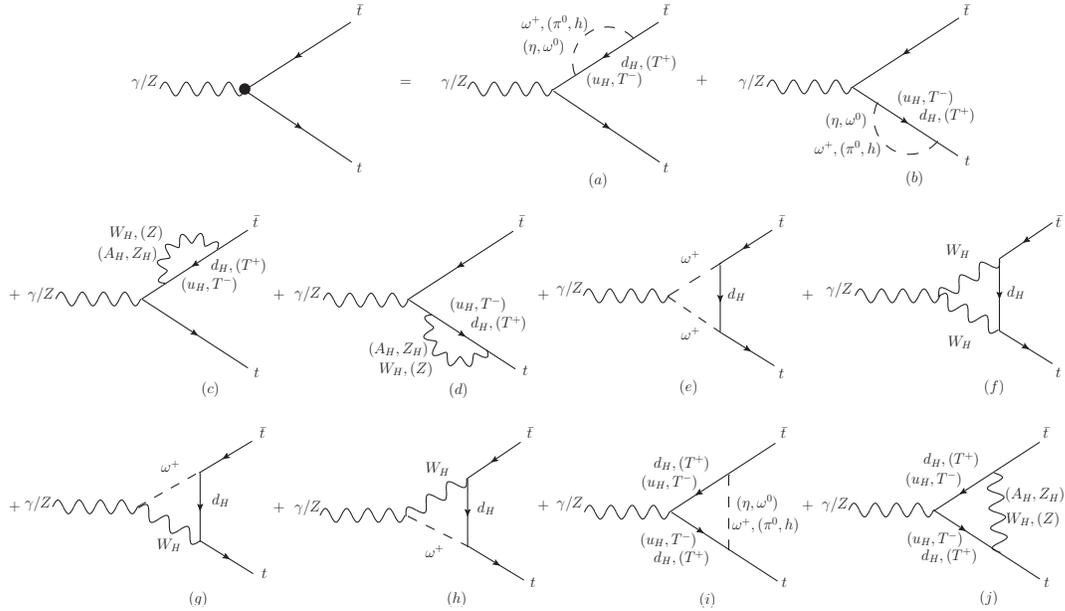}} \caption{Vertex correction
diagrams at one-loop level in the LHT model.}
\end{figure}
\begin{figure}[htbp]
\begin{center}
\scalebox{0.4}{\epsfig{file=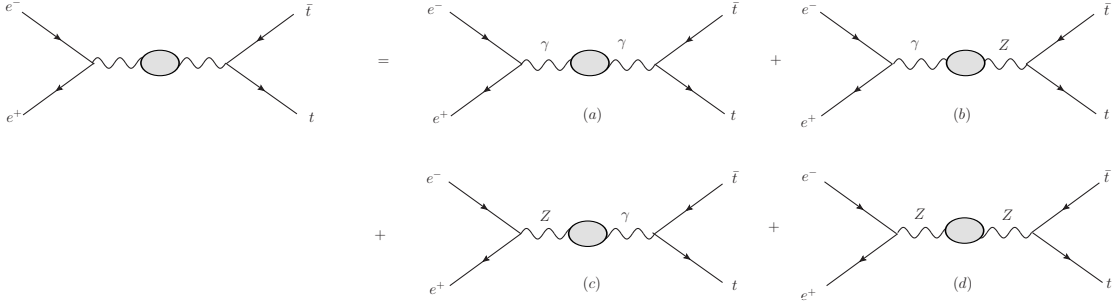}} \caption{Propagator
correction diagrams at one-loop level in the LHT model.}
\end{center}
\end{figure}
\begin{figure}[htbp]
\begin{center}
\scalebox{0.5}{\epsfig{file=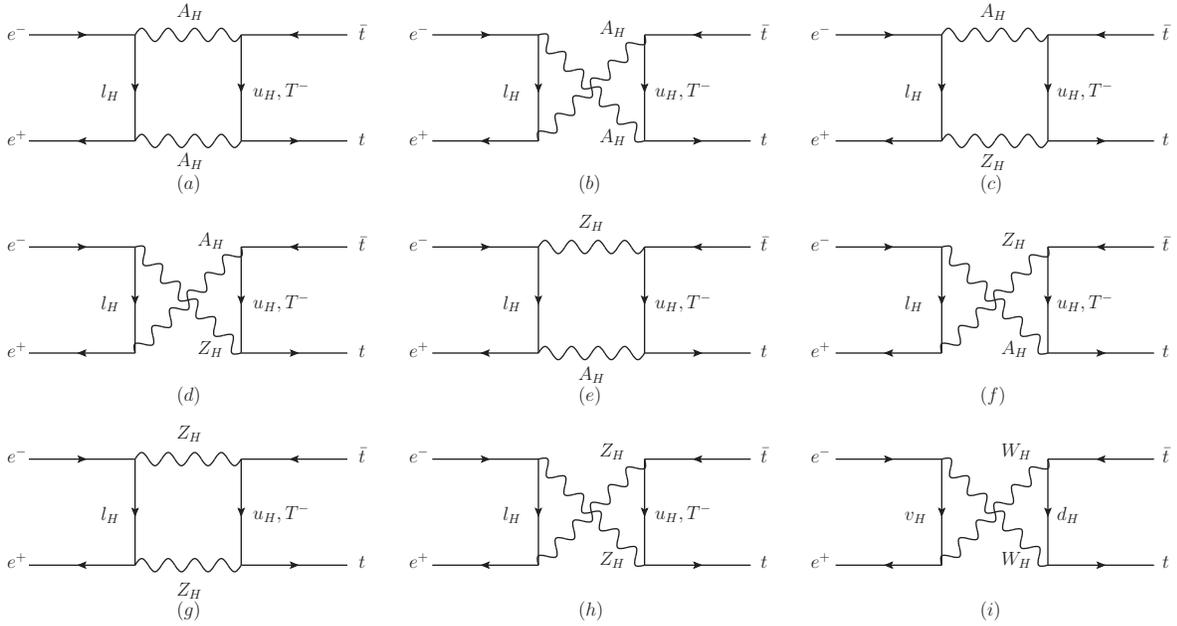}} \caption{Box diagrams at
one-loop level in the LHT model.}
\end{center}
\end{figure}

For the mirror fermion masses, we get $m_{F_{H}^{i}}$ at $\mathcal
O(\upsilon/f)$ and further assume
\begin{equation}
m_{F_{H}^{1}}=m_{F_{H}^{2}}=M_{12},m_{F_{H}^{3}}=M_{3}
\end{equation}

For the matrices $V_{Hu},V_{Hd}$ and $V_{Hl},V_{H\nu}$, considering
the constraints in Ref.\cite{15}, we follow Ref.\cite{16} to
consider the following two scenarios:

Scenario I: $V_{Hu}=1,V_{Hl}=V_{CKM}$

Scenario II: $V_{Hu}=1,V_{Hl}=V_{PMNS}$ \cite{17}
\begin{figure}[htbp]
\begin{center}
\scalebox{0.75}{\epsfig{file=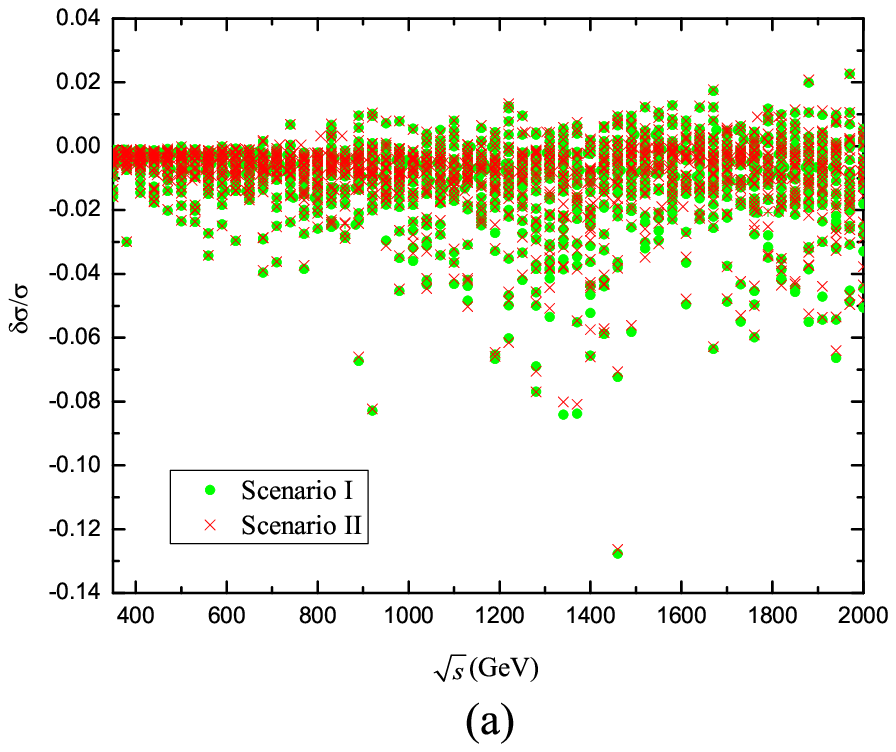}}\vspace{-1cm}
\hspace{-0.5cm} \scalebox{0.75}{\epsfig{file=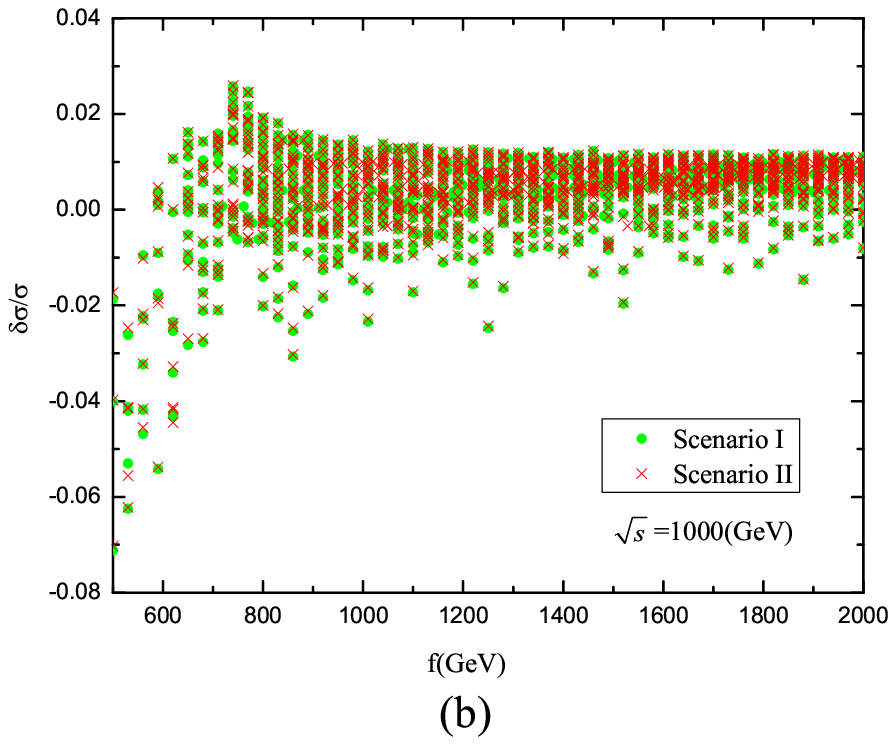}}
\caption{The relative correction of the production cross section
$\delta \sigma/\sigma$ as functions of the center-of-mass energy
$\sqrt{s}$ (a) and the scale $f$(b) in two scenarios, respectively.}
\end{center}
\end{figure}

In Fig.4(a), we discuss the dependance of $\delta \sigma/\sigma$ on
the center-of-mass energy $\sqrt{s}$ in two scenarios. Considering
the constraints of Ref.\cite{18}, we vary the parameters randomly as
follows: $M_{12}=300\sim3000GeV$, $M_{3}=300\sim3000GeV$,
$f=500\sim2000GeV$, $x_{L}=0.1\sim0.7$, and satisfy the relation
$m_{F_{H}^{i}}\leq 4.8f^{2}/TeV$. We can see $\delta \sigma/\sigma$
becomes larger with the $\sqrt{s}$ increasing. When the
center-of-mass energy $\sqrt{s}\geq600GeV$, some peaks appear duo to
the threshold effects of a pair of T-odd gauge bosons $W_{H}^{\pm}$
and the T-odd fermions. The maximum of the relative correction in
two scenarios can both reach about $-13\%$. \vspace{0.3cm}

In Fig.4(b), we discuss the dependance of $\delta \sigma/\sigma$ on
the scale $f$ in two scenarios. Considering the same constraints, we
vary the parameters randomly as follows: $M_{12}=300\sim3000GeV$,
$M_{3}=300\sim3000GeV$, $x_{L}=0.1\sim0.7$, and satisfy the relation
$m_{F_{H}^{i}}\leq 4.8f^{2}/TeV$. We can see $\delta \sigma/\sigma$
tends to zero with the $f$ increasing, which means that the
correction of the LHT model decouples with the $f$ increasing. The
threshold effects of a pair of T-odd particles still exist. The
maximum of the relative correction in two scenarios can both reach
about $-8\%$.
\begin{figure}[htbp]
\begin{center}
\scalebox{0.75}{\epsfig{file=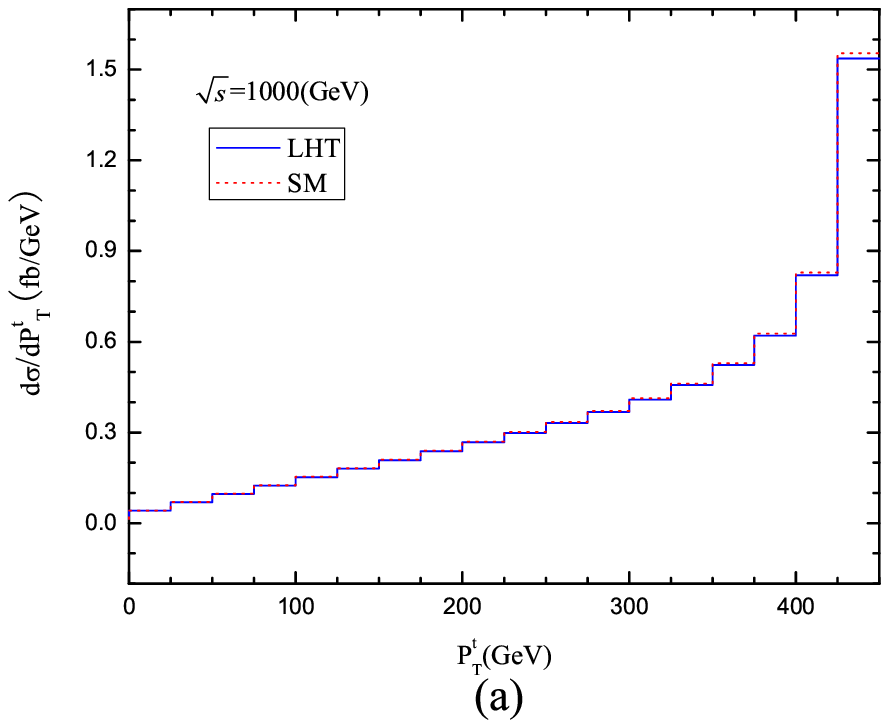}}\vspace{-1cm}
\hspace{-0.5cm} \scalebox{0.75}{\epsfig{file=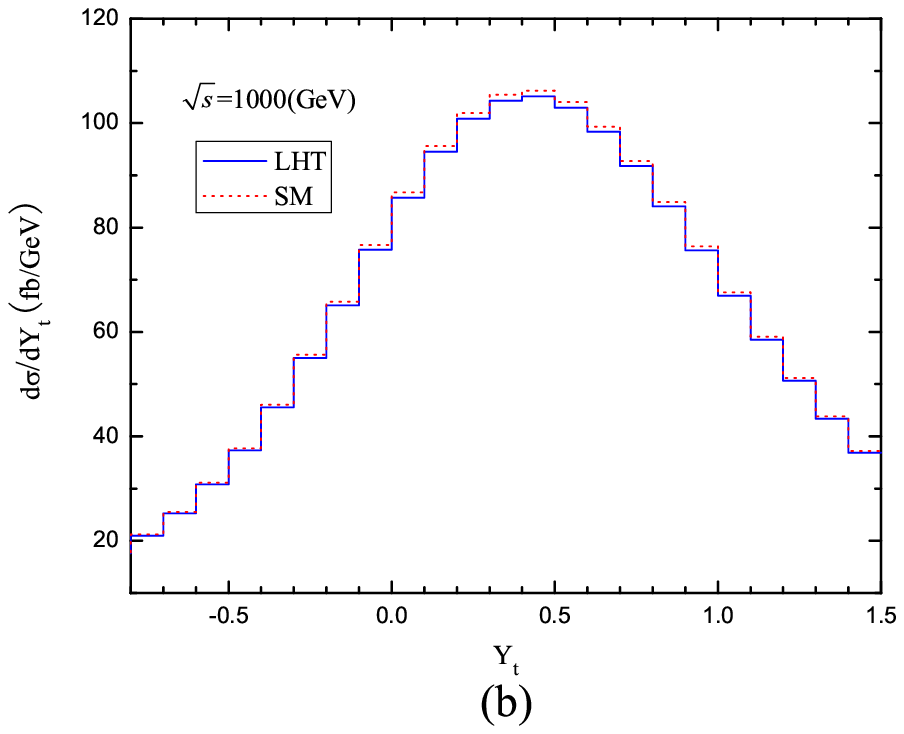}}
\caption{$d\sigma/dP_{T}^{t}$ as a function of the top-quark
transverse momentum $P_{T}^{t}$ (a) and $d\sigma/dY_{t}$ as a
function of the top-quark rapidity $Y_{t}$ (b)with
$\sqrt{s}=1000GeV$ in the SM and the LHT, respectively.}
\scalebox{0.75}{\epsfig{file=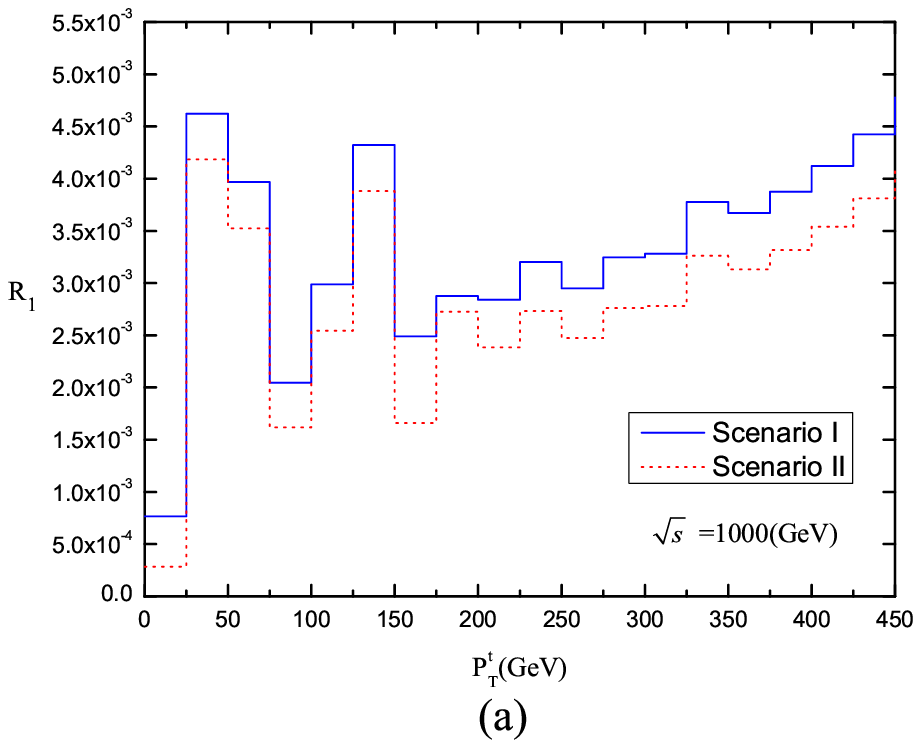}} \vspace{-1cm} \hspace{-0.5cm}
\scalebox{0.75}{\epsfig{file=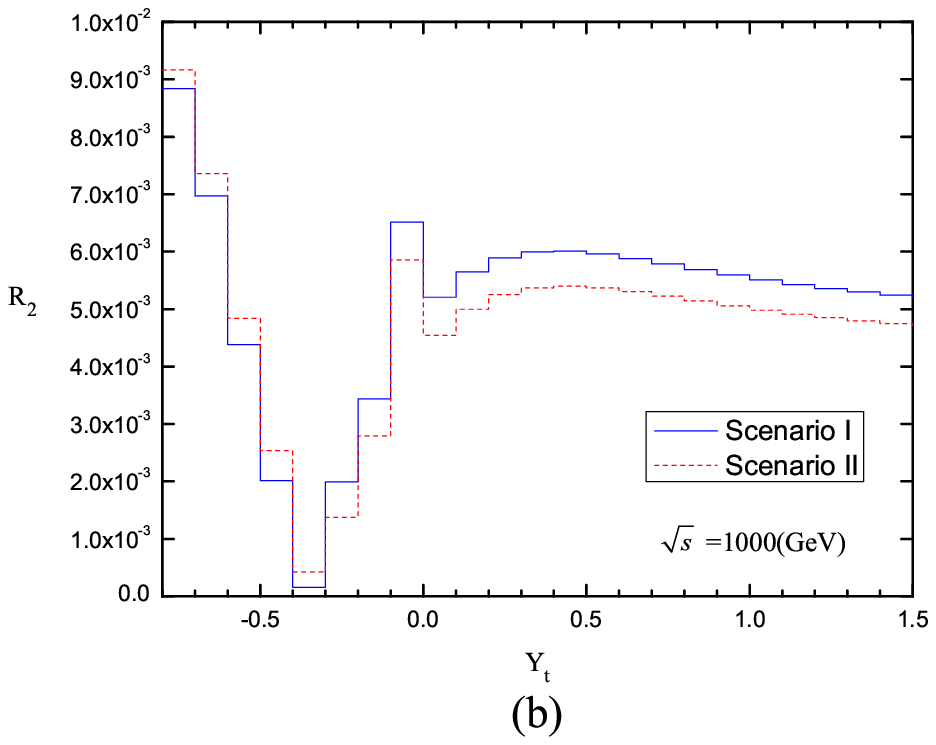}} \caption{$R_{1}$ as a function
of the top-quark transverse momentum $P_{T}^{t}$ (a) and $R_{2}$ as
a function of the top-quark rapidity $Y_{t}$ (b)with
$\sqrt{s}=1000GeV$ in two scenarios, respectively.}
\end{center}
\end{figure}

In Fig.5, we display the transverse momentum distribution and the
rapidity of the final state top quark for $M_{12}=600GeV,
M_{3}=1000GeV, f=1000GeV,x_{L}=0.4,\sqrt{s}=1000GeV$ in the LHT
model and the SM, respectively. Because the
$d\sigma_{tot}/dP_{T}^{t}$ and the $d\sigma_{tot}/dY_{t}$ change
very little in two scenarios, we don't distinguish them and only
show one plot as example.

From Fig.5(a), we can see that the transverse momentum distribution
behaviour in the LHT model is similar as the SM and the values of
the SM are slightly larger than the values of the LHT model. The
transverse momentum values of top-quark ranging from 300 to 450
$GeV$ make the main contribution to the $d\sigma/dP_{T}^{t}$, which
is more significant in the regions around $P_{T}^{t}\sim $ 425$GeV$
than in other regions.

From Fig.5(b), same as above, we can see the rapidity behaviour in
the LHT model is similar as the SM and the values of the SM are
slightly larger than the values of the LHT model. The rapidity
values of top quark ranging from 0 to 0.9 make the main contribution
to the $d\sigma/dY_{t}$, which is more significant in the regions
around $Y_{t}\sim 0.4$ than in other regions. The rapidity
distribution is asymmetric at the zero rapidity, which is caused by
the $Z$ boson mediated in the $S$-channel of the process
$e^{+}e^{-}\rightarrow t\bar{t}$.

In Fig.6, we display the relative deviations of the transverse
momentum distribution and the rapidity of the final state top quark
from the SM for $M_{12}=600GeV, M_{3}=1000GeV,
f=1000GeV,x_{L}=0.4,\sqrt{s}=1000GeV$ in two scenarios,
respectively. The relative ratio $R_{1}(P_{T}^{t})$ for the
top-quark transverse momentum distribution $(P_{T}^{t})$ and the
relative ratio $R_{2}(Y_{t})$ for the top-quark rapidity $(Y_{t})$
of the process $e^{+}e^{-}\rightarrow t\bar{t}$ can be defined as
\begin{eqnarray}
R_{1}&=&\left|\frac{d\sigma_{tot}/dP_{T}^{t}-d\sigma_{SM}/dP_{T}^{t}}{d\sigma_{SM}/dP_{T}^{t}}\right|\\
R_{2}&=&\left|\frac{d\sigma_{tot}/dY_{t}-d\sigma_{SM}/dY_{t}}{d\sigma_{SM}/dY_{t}}\right|
\end{eqnarray}

From Fig.6(a), we can see that is more significant in the regions
around $P_{T}^{t}\sim $ 25$GeV,125GeV$ and $425GeV$ than in other
regions. Furthermore, we can see $R_{1}$ is larger in scenario I
than in scenario II. \vspace{0.3cm}

From Fig.6(b), we can see that is more significant in the regions
around $Y_{t}\sim -0.8$ than in other regions. The rapidity
distribution is asymmetric at the zero rapidity, which is caused by
the $Z$ boson mediated in the $S$-channel of the process
$e^{+}e^{-}\rightarrow t\bar{t}$. Different from $R_{1}$, we can see
$R_{2}$ is larger in scenario I when $Y_{t}>-0.3$ but smaller in
scenario I when $Y_{t}<-0.3$.

\section{Conclusions} \noindent In the framework of the LHT model, we studied the
one-loop contributions of the T-odd particles to the process
$e^{+}e^{-}\rightarrow t\bar{t}$ for two different scenarios. The
ILC is designed with the center-of-mass energy $\sqrt{s}$=$300$
${\sim}$ $1500GeV$ and a precision of around 5\% can be reached with
$\sqrt{s}$=$800\sim 1000GeV$ and the integrated luminosity $\mathcal
L_{int}\simeq 1000fb^{-1}$\cite{19}. The relative correction of the
cross section in the LHT model is significant so that the possible
signals of the LHT model might be observed at the ILC. This is
really interesting in testing the SM and probing the NP.
\vspace{1.5cm}\\
\textbf{Acknowledgments}\\
We thank Junjie Cao for providing the calculation programs and thank
Lei Wu for useful discussions and suggestions. This work is
supported by the National Natural Science Foundation of China under
Grant Nos.10775039, 11075045, by Specialized Research Fund for the
Doctoral Program of Higher Education under Grant No.20094104110001
and by HASTIT under Grant No.2009HASTIT004.
%\begin{appendix}
%\documentcalss[fleqn]{article}
%\newpage
\begin{center}
\textbf{Appendix: The expression of the renormalization vertex
$\hat{\Gamma}^{\mu}_{Vt\bar{t}}(V=\gamma,Z)$ and the renormalization
propagator $-i\hat{\Sigma}^{V_{1}V_{2}}_{\mu\nu}$} \cite{20}
\end{center}
(I)Renormalization vertex
\begin{figure}[htbp]
\scalebox{0.5}{\epsfig{file=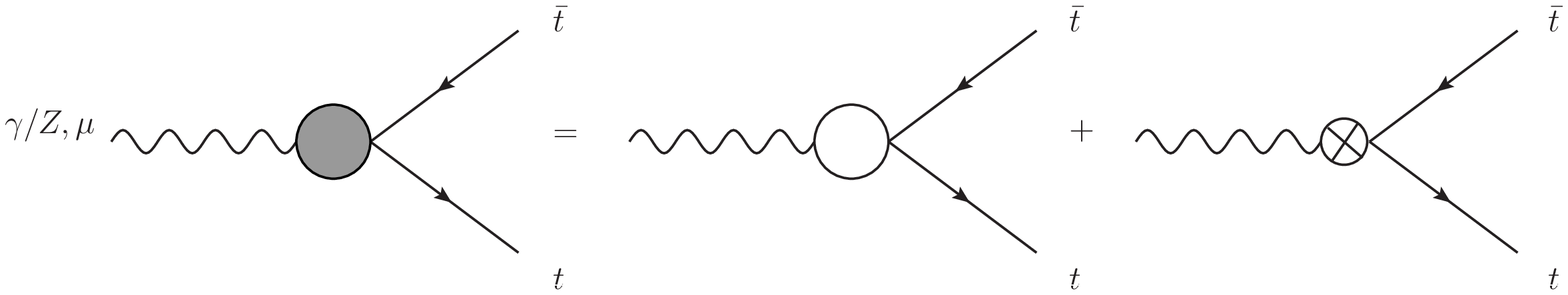}}
\end{figure}
\begin{eqnarray}
\hat{\Gamma}^{\mu}_{\gamma t\bar{t}}&=&\Gamma^{\mu}_{\gamma
t\bar{t}}-ieQ_{t}\gamma^{\mu}(\delta Z_{V}^{t}-\gamma_{5}\delta
Z_{A}^{t}-\frac{S_{W}}{2C_{W}}\delta Z_{ZA})
+ie\gamma^{\mu}(v_{t}-a_{t}\gamma_{5})\frac{1}{2}\delta
Z_{ZA}\nonumber\\
\hat{\Gamma}^{\mu}_{Zt\bar{t}}&=&\Gamma^{\mu}_{Zt\bar{t}}-ie\gamma^{\mu}(v_{t}-a_{t}\gamma_{5})\frac{C_{W}}{2S_{W}}
\delta Z_{ZA}-ieQ_{t}\gamma^{\mu}\frac{1}{2} \delta
Z_{ZA}\nonumber\\
&+&ie\gamma^{\mu}(v_{t}-a_{t}\gamma_{5})\delta
Z_{V}^{t}-ie\gamma^{\mu}\gamma_{5}(v_{t}-a_{t}\gamma_{5})\delta
Z_{A}^{t}\nonumber
\end{eqnarray}

where
\begin{eqnarray*}
v_{t}\equiv\frac{I_{t}^{3}-2Q_{t}S_{W}^{2}}{2C_{W}S_{W}},\quad
a_{t}\equiv\frac{I_{t}^{3}}{2C_{W}S_{W}},
\quad I_{t}^{3}=\frac{1}{2},\quad Q_{t}=\frac{2}{3}~~~~~~~~~~~\\
\quad \delta Z_{ZA}=2\frac{\Sigma_{T}^{AZ}(0)}{M_{Z_{L}}^{2}}~~~~~~~~~~~~~~~~~~~~~~~~~~~~~~~~~~~~~~~~~~~~~~~~~~~~~~~~~~~~~~~~~~\\
\delta
Z_{L}^{t}=Re\Sigma_{L}^{t}(m_{t}^{2})+m_{t}^{2}\frac{\partial}{\partial
P_{t}^{2}}Re[\Sigma_{L}^{t}(P_{t}^{2})+\Sigma_{R}^{t}(P_{t}^{2})+2\Sigma_{S}^{t}(P_{t}^{2})]|_{P_{t}^{2}=m_{t}^{2}}\\
\delta
Z_{R}^{t}=Re\Sigma_{R}^{t}(m_{t}^{2})+m_{t}^{2}\frac{\partial}{\partial
P_{t}^{2}}Re[\Sigma_{L}^{t}(P_{t}^{2})+\Sigma_{R}^{t}(P_{t}^{2})+2\Sigma_{S}^{t}(P_{t}^{2})]|_{P_{t}^{2}=m_{t}^{2}}\\
\delta Z_{V}^{t}=\frac{1}{2}(\delta Z_{L}^{t}+\delta
Z_{R}^{t}),\delta Z_{A}^{t}=\frac{1}{2}(\delta Z_{L}^{t}-\delta
Z_{R}^{t})~~~~~~~~~~~~~~~~~~~~~~~~~~~~~~~
\end{eqnarray*}
\begin{eqnarray*}
\hat{\Gamma}^{LHT,\mu}_{\gamma t\bar{t}}&=&\Gamma^{\mu}_{\gamma
t\bar{t}}(\eta)+ \Gamma^{\mu}_{\gamma
t\bar{t}}(\omega^{0})+\Gamma^{\mu}_{\gamma t\bar{t}}(\omega^{\pm})+
\Gamma^{\mu}_{\gamma t\bar{t}}(A_{H})+\Gamma^{\mu}_{\gamma
t\bar{t}}(Z_{H})+
\Gamma^{\mu}_{\gamma t\bar{t}}(W_{H}^{\pm})+\Gamma^{\mu}_{\gamma t\bar{t}}(\omega^{\pm},W_{H}^{\pm})\\
&+&\delta\Gamma^{\mu}_{\gamma t\bar{t}}(\eta)+
\delta\Gamma^{\mu}_{\gamma
t\bar{t}}(\omega^{0})+\delta\Gamma^{\mu}_{\gamma
t\bar{t}}(\omega^{\pm})+ \delta\Gamma^{\mu}_{\gamma
t\bar{t}}(A_{H})+\delta\Gamma^{\mu}_{\gamma t\bar{t}}(Z_{H})+
\delta\Gamma^{\mu}_{\gamma t\bar{t}}(W_{H}^{\pm})\\
\hat{\Gamma}^{LHT,\mu}_{Zt\bar{t}}&=&\Gamma^{\mu}_{Zt\bar{t}}(\eta)+
\Gamma^{\mu}_{Zt\bar{t}}(\omega^{0})+\Gamma^{\mu}_{Zt\bar{t}}(\omega^{\pm})+
\Gamma^{\mu}_{Zt\bar{t}}(A_{H})+\Gamma^{\mu}_{Zt\bar{t}}(Z_{H})+
\Gamma^{\mu}_{Zt\bar{t}}(W_{H}^{\pm})+\Gamma^{\mu}_{Zt\bar{t}}(\omega^{\pm},W_{H}^{\pm})\\
&+&\delta\Gamma^{\mu}_{Zt\bar{t}}(\eta)+
\delta\Gamma^{\mu}_{Zt\bar{t}}(\omega^{0})+\delta\Gamma^{\mu}_{Zt\bar{t}}(\omega^{\pm})+
\delta\Gamma^{\mu}_{Zt\bar{t}}(A_{H})+\delta\Gamma^{\mu}_{Zt\bar{t}}(Z_{H})+
\delta\Gamma^{\mu}_{Zt\bar{t}}(W_{H}^{\pm})
\end{eqnarray*}
\newpage
(II)Renormalization propagator
\begin{figure}[htbp]
\scalebox{0.5}{\epsfig{file=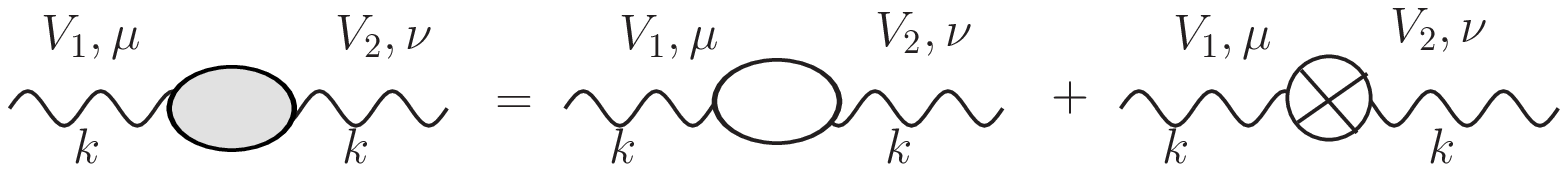}}
\end{figure}
\begin{eqnarray*}
-i\hat{\Sigma}^{V_{1}V_{2}}_{\mu\nu}(k)=-i\Sigma^{V_{1}V_{2}}_{\mu\nu}(k)+(-i\delta\Sigma^{V_{1}V_{2}}_{\mu\nu}(k))
\end{eqnarray*}
where
\begin{eqnarray*}
\Sigma^{V_{1}V_{2}}_{\mu\nu}(k)&=&g_{\mu\nu}\Sigma^{V_{1}V_{2}}_{T}(k)+k_{\mu}k_{\nu}\Sigma^{V_{1}V_{2}}_{L}(k)\\
V_{1},V_{2}&=&A,Z\\
\delta\Sigma^{AA}_{\mu\nu}(k)&=&g_{\mu\nu}[\delta Z_{AA}k^{2}]\\
\delta\Sigma^{AZ}_{\mu\nu}(k)&=&g_{\mu\nu}[\frac{1}{2}(k^{2}-M^{2})\delta
Z_{ZA}+\frac{1}{2}\delta Z_{AZ}k^{2}]\\
\delta\Sigma^{ZZ}_{\mu\nu}(k)&=&g_{\mu\nu}[-\delta
M^{2}_{Z}+(k^{2}-M^{2})\delta Z_{ZZ}]\\
\delta M^{2}_{Z}&=&Re\Sigma^{ZZ}_{T}(M^{2}_{Z})\\
\delta Z_{ZZ}&=&-Re\frac{\partial\Sigma^{ZZ}_{T}(k^{2})}{\partial
k^{2}}|_{k^{2}=M_{Z}^{2}},~~~~\delta
Z_{AZ}=-2\frac{Re\Sigma^{AZ}_{T}(M^{2})}{M^{2}} \\
\delta
Z_{ZA}&=&2\frac{\Sigma_{T}^{AZ}(0)}{M_{Z_{L}}^{2}},~~~~~~~~~~~~~~~~~~~~\delta
Z_{AA}=-\frac{\partial\Sigma^{AA}_{T}(k^{2})}{\partial
k^{2}}|_{k^{2}=0}
\end{eqnarray*}
\begin{eqnarray*}
\hat{\Sigma}^{\gamma\gamma}_{\mu\nu}&=&\Sigma^{\gamma\gamma}_{\mu\nu}(f\bar{f})+\Sigma^{\gamma\gamma}_{\mu\nu}(W_{H}^{\pm})+\Sigma^{\gamma\gamma}_{\mu\nu}(W_{H}^{\pm},\omega^{\pm})
+\Sigma^{\gamma\gamma}_{\mu\nu}(\omega^{\pm})+\Sigma^{\gamma\gamma}_{\mu\nu}(u^{\pm})\\&+&\delta\Sigma^{\gamma
\gamma}_{\mu\nu}(f\bar{f})+\delta\Sigma^{\gamma\gamma}_{\mu\nu}(W_{H}^{\pm})+\delta\Sigma^{\gamma\gamma}_{\mu\nu}(W_{H}^{\pm},\omega^{\pm})
+\delta\Sigma^{\gamma\gamma}_{\mu\nu}(\omega^{\pm})+\delta\Sigma^{\gamma\gamma}_{\mu\nu}(u^{\pm})\\
\hat{\Sigma}^{\gamma Z}_{\mu\nu}&=&\Sigma^{\gamma
Z}_{\mu\nu}(f\bar{f})+\Sigma^{\gamma
Z}_{\mu\nu}(W_{H}^{\pm})+\Sigma^{\gamma
Z}_{\mu\nu}(W_{H}^{\pm},\omega^{\pm}) +\Sigma^{\gamma
Z}_{\mu\nu}(\omega^{\pm})+\Sigma^{\gamma
Z}_{\mu\nu}(u^{\pm})\\&+&\delta\Sigma^{\gamma
Z}_{\mu\nu}(f\bar{f})+\delta\Sigma^{\gamma
Z}_{\mu\nu}(W_{H}^{\pm})+\delta\Sigma^{\gamma
Z}_{\mu\nu}(W_{H}^{\pm},\omega^{\pm})
+\delta\Sigma^{\gamma Z}_{\mu\nu}(\omega^{\pm})+\delta\Sigma^{\gamma Z}_{\mu\nu}(u^{\pm})\\
\hat{\Sigma}^{ZZ}_{\mu\nu}&=&\Sigma^{ZZ}_{\mu\nu}(f\bar{f})+\Sigma^{ZZ}_{\mu\nu}(W_{H}^{\pm})+\Sigma^{ZZ}_{\mu\nu}(W_{H}^{\pm},\omega^{\pm})
+\Sigma^{ZZ}_{\mu\nu}(\omega^{\pm})\\&+&\Sigma^{ZZ}_{\mu\nu}(Z_{H},\eta)+\Sigma^{ZZ}_{\mu\nu}(\omega^{0},\eta)+\Sigma^{ZZ}_{\mu\nu}(\eta)+\Sigma^{ZZ}_{\mu\nu}(\omega^{0})+\Sigma^{ZZ}_{\mu\nu}(u^{\pm})\\&+&\delta\Sigma^{Z
Z}_{\mu\nu}(f\bar{f})+\delta\Sigma^{ZZ}_{\mu\nu}(W_{H}^{\pm})+\delta\Sigma^{\gamma\gamma}_{\mu\nu}(W_{H}^{\pm},\omega^{\pm})
+\delta\Sigma^{ZZ}_{\mu\nu}(\omega^{\pm})\\&+&\delta\Sigma^{ZZ}_{\mu\nu}(Z_{H},\eta)+\delta\Sigma^{ZZ}_{\mu\nu}(\omega^{0},\eta)+\delta\Sigma^{ZZ}_{\mu\nu}(\eta)+\delta\Sigma^{ZZ}_{\mu\nu}(\omega^{0})+\delta\Sigma^{ZZ}_{\mu\nu}(u^{\pm})\\
\end{eqnarray*}
%\end{appendix}

\end{document}